\documentclass[12pt]{iopart}
\usepackage{graphicx}
\usepackage{latexsym}
\usepackage{amsthm}

\usepackage{iopams}
\usepackage{url}
\def\apj{ApJ\,  }
\def\apjl{ApJ\,  }
\def\araa{ARA\&A  }

\def\mnras{MNRAS\,  }

\def\nat{Nature\,  }
\def\pasj{PASJ\,  }
\def\pasp{PASP  }

\def\sun{\hbox{$\odot$}}
\def\h0units{\mathrm{km\,s^{-1}\,Mpc^{-1}}}
\def\cunits{\mathrm{km\,s^{-1}}}

\newcommand{\om}{\Omega_{\rm M}}

\newcommand{\ola}{\Omega_{\Lambda}}
\newcommand{\dl}{d_{\rm{L}}}
\newcommand{\dldieci}{d_{\rm{L,10}}}
\newcommand{\da}{d_{\rm{A}}}
\newcommand{\dadieci}{d_{\rm{A,10}}}
\newtheorem{mydeftheorem}{Theorem}
\def\apj{ApJ\,  }
\def\apjl{ApJ\,  }
\def\araa{ARA\&A  }

\def\mnras{MNRAS\,  }

\def\nat{Nature\,  }
\def\pasj{PASJ\,  }
\def\pasp{PASP  }

\begin{document}
\title
{
The Ring Produced by an   
Extra-Galactic Superbubble
in Flat Cosmology
}
\vskip  1cm
\author     {L. Zaninetti}
\address    {
Physics Department,
 via P.Giuria 1,\\ I-10125 Turin,Italy 
}
\ead {zaninetti@ph.unito.it}

\begin {abstract}
A superbubble which advances in a symmetric
Navarro--Frenk--White  
density profile 
or in an auto-gravitating density profile
generates a thick shell 
 with a radius that can reach 10 kpc.
The application of the symmetric and asymmetric image theory
to this thick 3D shell produces a ring in 
the 2D map of intensity and a characteristic `U' shape
in the case of 1D cut of the  intensity.
A comparison of such a ring originating from a superbubble
is made with the Einstein's ring.
A Taylor approximation of order 10 
for the angular diameter distance
is derived  in order to deal 
with high values of the redshift. 
\end{abstract}
\vspace{2pc}
\maketitle 

\section{Introduction}

A first  theoretical prediction of the existence of gravitational
lenses (GL) is due to  Einstein
\cite{Einstein1936} where the 
formulae for the optical properties of a gravitational lens
for star A and B were derived.
A first sketch  which dates back to 1912 is reported 
at  p.585 in  \cite{Einstein_3_1994}.
The  historical context of the GL is outlined in
 \cite{Valls-Gabaud2006} 
and ONLINE information can be found  at    
\url{http://www.einstein-online.info/}. 
After  43 years  a first GL  was observed in the form of 
a close pair of blue stellar objects of magnitude 17
with a separation of 5.7 arc sec at redshift 1.405,
 0957 + 561 A, B,   see \cite{Walsh1979}.
This double system is also known as the "Twin Quasar"
and a 
Figure  which reports a 2014 image  
of  the  Hubble Space Telescope (HST) for   
objects A and B is available at
\url{https://www.nasa.gov/content/goddard/hubble-hubble-seeing-double/}.  

At the moment of writing, the GL is used 
routinely as an explanation 
for lensed objects, see the following reviews 
\cite{Blandford1992,Mellier1999,Refregier2003,Treu2010}.
As an example of current observations, 
28 gravitationally lensed quasars have been observed 
by the Subaru Telescope, see \cite{Rusu2016},
where for   each system  a  mass model was derived.
Another example is given 
by the SDSS-III Baryon Oscillation Spectroscopic 
Survey (BOSS), see
\cite{More2016},
where 13  two-image quasar lenses  have been observed
and the relative Einstein's radius reported in arcsec.
A first classification separates the {\it strong lensing}, 
such as an Einstein ring (ER)  and the arcs from the {\it weak lensing},
such as the shape deformation of background galaxies.

The {\it strong lensing} is verified 
when the light from a distant background source, 
such as a galaxy or quasar, is
deflected into multiple paths by an intervening galaxy or a cluster of
galaxies  producing multiple images of the background source:
examples are the ER   and the multiple  arcs in cluster
of galaxies, see \url{https://apod.nasa.gov/apod/ap160828.html}.

In the case of  {\it weak lensing}  
the lens is not strong enough to form multiple images or arcs,
but the source can be distorted: both stretched (shear) and
magnified (convergence), see \cite{Kilbinger2015}
and  \cite{DePaolis2016}.
The first cluster of galaxies 
observed with the {\it weak lensing}   effect  
is reported by \cite{Wittman2001}.

We now introduce supershells, which were unknown 
when the GL was postulated. 
Supershells  started to be observed 
{\it firstly} 
in our galaxy 
by \cite{heiles1979},
where 17 expanding H I shells were classified, 
and  {\it secondly}  
in external galaxies, see as an example
\cite{Sanchez2015}, where many  supershells were observed
in NGC 1569.
In order to model such complex objects, 
the term super bubble (SB) has  been introduced but
unfortunately the astronomers often 
associate the SBs  with sizes of $\approx$ 10--100 pc
and the 
supershells
with  ring-like structures with sizes of $\approx$  1 kpc.
At the same time an application of the theory of image
explains  
the  limb-brightening 
visible  on the maps of intensity 
of SBs and allows associating the observed 
filaments to undetectable SBs,
see \cite{Zaninetti2012g}.

This paper derives, in Section \ref{secflatcosmo}, an 
approximate solution for the 
angular diameter distance in flat cosmology.
Section \ref{secer} briefly reviews the existing knowledge 
of ERs.
Section \ref{secmotion} derives an equation of motion for an SB 
in a Navarro--Frenk--White (NFW) density profile.
Section \ref{secmotionasymm} adopts a
recursive equation in order to model
an asymmetric motion for an SB 
in an auto-gravitating density profile.

Section \ref{ringsec} applies the symmetrical and 
the asymmetrical 
image theory 
to the advancing shell of an SB.

\section{The flat  cosmology}

\label{secflatcosmo}

\begin{figure*}
\begin{center}
\includegraphics[width=10cm]{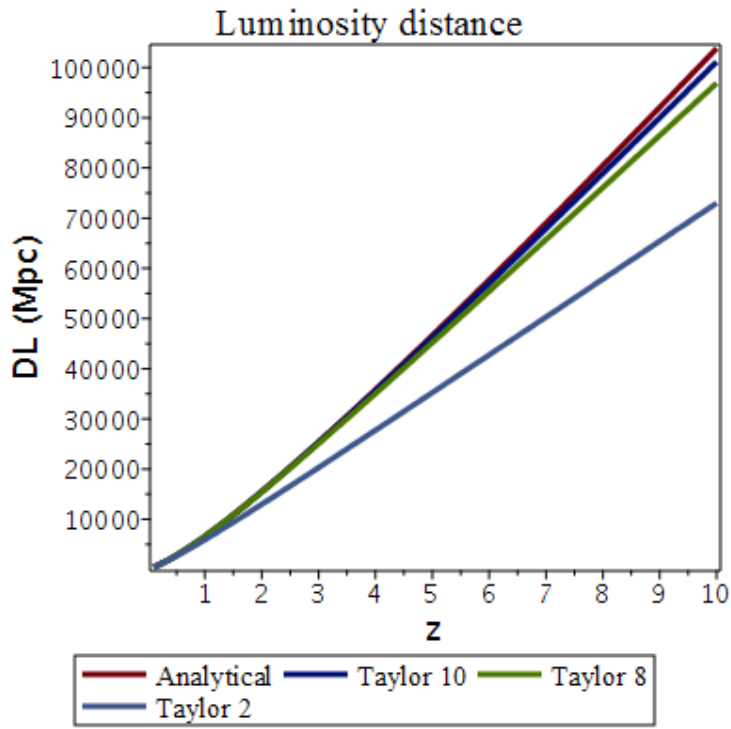}
\end {center}
\caption
{
The analytical solution (red line ) for the luminosity distance  
and three Taylor approximated solutions
with color coding as in the legend.
}
\label{foursolutions}
    \end{figure*}

Following eq. (2.1) of \cite{Adachi2012},
the luminosity distance in flat cosmology, 
$\dl$, is
\begin{equation}
  \dl(z;c,H_0,\om) = \frac{c}{H_0} (1+z) \int_{\frac{1}{1+z}}^1
  \frac{da}{\sqrt{\om a + (1-\om) a^4}} \quad ,
  \label{lumdistflat}
\end{equation}
where $H_0$
is the Hubble constant expressed in     $\h0units$,
$c$ is the velocity  of light expressed in $\cunits$,
$z$ is the redshift,
$a$ is the scale factor,
and  $\om$ is
\begin{equation}
\om = \frac{8\pi\,G\,\rho_0}{3\,H_0^2}
\quad ,
\end{equation}
where $G$ is the Newtonian gravitational constant and
$\rho_0$ is the mass density at the present time.
An analytical solution for the luminosity  
distance  exists
in the complex plane, see \cite{Zaninetti2015b}.
Here we deal with  an  approximate  solution
for the luminosity distance 
in the framework of a  flat  universe 
adopting the same cosmological parameters of 
 \cite{Tamura2015} which are
$H_0=72\,\h0units$, $\om=0.26$ and $\ola=0.74$.
An  approximate solution for the 
luminosity distance, $\dldieci (z)$,   is given
by a Taylor   expansion 
of order 10
about $a=1$  for  
the argument of the integral (\ref{lumdistflat})
\begin{eqnarray}
\dldieci (z) =
4163.78\, \left( 1+z \right)  \Bigg (  2.75+ 0.12882\,
 \left( 1+z \right) ^{-10}
 \nonumber \\
- 1.34123\, \left( 1+z \right) ^{-9}+
 6.23877\, \left( 1+z \right) ^{-8}- 17.0003\, \left( 1+z
 \right) ^{-7}
\nonumber  \\
+ 29.8761\, \left( 1+z \right) ^{-6}- 35.1727\,
 \left( 1+z \right) ^{-5}+ 28.2558\, \left( 1+z \right) ^{-4}
\nonumber \\
-
 16.5327\, \left( 1+z \right) ^{-3}+ 9.26107\, \left( 1+z
 \right) ^{-2}- 6.46\, \left( 1+z \right) ^{-1} \Bigg ) 
\quad .
\end{eqnarray}
More details  on the analytical solution for the luminosity distance 
in the case  of  flat cosmology can be found 
in \cite{Zaninetti2016b}
and  
Figure \ref{foursolutions}
reports the comparison between the above analytical solution,
and  Taylor   expansion 
of order 10, 8 and 2.

The    goodness of the Taylor  approximation 
is evaluated
through the percentage error, $\delta$,
which is
\begin{equation}
\delta = \frac{\big | \dl(z) - \dldieci(z)_{10} \big |}
{\dl(z) } \times 100
\quad .
\end{equation}
As an example, Table  \ref{tablepercentage} reports 
the  percentage  error  at z= 4 for three order
of expansion; is clear the progressive 
decrease of the percentage  error  with  the increase 
in the order of expansion.

\begin{table}[ht!]
\caption {
Percentage error between analytical solution and 
approximated Taylor 
solution of a given order at z=4.
}
\label{tablepercentage}
\begin{center}
\begin{tabular}{|c|c|}
\hline
Taylor ~order          & $\delta$  \\
\hline                                                   
2      &  22.6 \%  \\
8      &  2.2  \%  \\
10     &  0.61 \%  \\
\hline
\end{tabular}
\end{center}
\end{table}

Another   useful distance is the angular 
diameter distance, $\da$,
which is 
\begin{equation}
\da = \frac{\dl}{(1+z)^2}
\quad ,
\end{equation}
see \cite{Etherington1933} and
the Taylor approximation for the angular diameter distance, 
$\dadieci$ 
\begin{equation}
\dadieci = \frac{\dldieci}{(1+z)^2}
\quad .
\label{angulardistance10}
\end{equation}

As a practical  example of the above  equation, 
the angular scale of $1\,arcsec$ is
$7.73$\ kpc  at $z=3.042$
when  \cite{Tamura2015} quotes $7.78$\ kpc:
this  means  a percentage error  of 0.63\%
between the two values.
Another check can be done with the
Ned Wright's Cosmology Calculator \cite{Wright2006} 
available at \url{http://www.astro.ucla.edu/~wright/CosmoCalc.html}:
it quotes a scale of 7.775 $kpc\,arcsec^{-1}$
which means  a percentage error  of 0.57\%
with respect to our value.
In this section we have derived the cosmological scaling 
that allows to fix the dimension of the ER.
  
\section{The ER}

\label{secer}

This section reviews the simplest version of the ER 
and reports the observations of two recent ERs.

\subsection{The theory }

In the case of a circularly  symmetric   
lens and when the source 
and the length are on the same line of sight, 
the ER  radius in radiant is  
\begin{equation}
\theta_E = \sqrt 
{ 
\frac{ 4 G M (\theta_E)}{c^2}
\frac{D_{ds}}{D_d \,D_s} 
}  
\quad ,
\end{equation}
where $M (\theta_E)$ is the mass enclosed 
inside the ER radius,
$D_{d,s,ds}$ are 
the lens, 
source and lens--source distances, respectively,
$G$ is the Newtonian gravitational constant, 
and $c$ is the velocity of light, 
see eq.~(20) in \cite{Narayan1996} 
and eq.~(1)  in \cite{Bettinelli2016}.
The mass of the ER 
can be expressed in 
units of solar mass, $M_{\sun}$:
\begin{equation}
M (\theta_{E,arcsec}) =
1.228 \, 10^8 
\,{\frac {{\Theta_{{{\it E, arcsec}}}}^{2}
D_{{{\it ds},{\mathrm 
{Mpc}}}}D_{{s,{\mathrm {Mpc}}}}}{D_{{d,{\mathrm {Mpc}}}}}}
\, M_{\sun}
\quad , 
\label{massasun}
\end{equation}
  
where $\theta_{E,arcsec}$ is the 
ER  radius in ${arcsec}$
and the three distances are expressed in Mpc.

\subsection{The galaxy--galaxy lensing system SDP.81}

\label{secsdp81} 
\begin{figure*}
\begin{center}
\includegraphics[width=10cm]{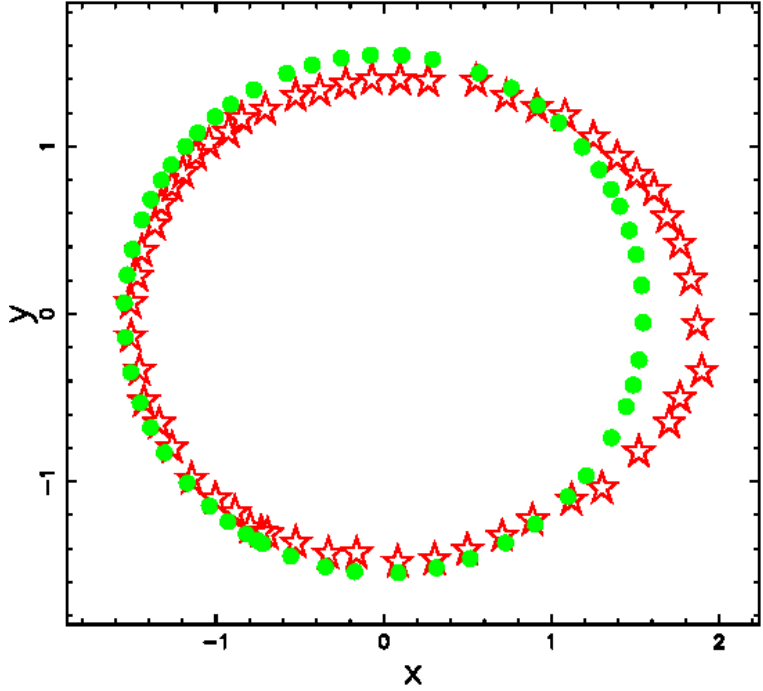}
\end {center}
\caption
{
Real data of SDP.81 ring (empty red stars) 
and averaged circle      (full  green points).
The real data are extracted by the author from Figure 6
in   \cite{Hezaveh2016}
using  WebPlotDigitizer. 
}
\label{real_ring}
    \end{figure*}

The ring associated with the galaxy SDP.81,
see \cite{Eales2010}, 
 is  generally explained by a GL.
In this framework we have a foreground galaxy at 
$z=0.2999$ and a background galaxy at $z=0.3042$.
This ring  has  been studied with the 
Atacama Large Millimeter/sub-millimeter Array (ALMA)
by 
\cite{Tamura2015,ALMA2015,Rybak2015,Hatsukade2015,Wong2015,Hezaveh2016}.
The system SDP.81 
as analysed 
by  ALMA  presents 14  molecular clumps along the two main
lensed arcs.
We can therefore speak of the ring appearance as 
a `grand design' and we now test the circular hypothesis.
In order to  test the departure from a circle,  an
observational percentage  of
reliability is  introduced that uses
both the size and the shape,
\begin{equation}
\epsilon_{\mathrm {obs}} =100(1-\frac{\sum_j |R_{\mathrm {obs}}-R_{\mathrm
{ave}}|_j}{\sum_j
{R_{\mathrm {obs}}}_j})
\label{reliabilitypercentage}
\quad , 
\end{equation}
where $R_{\mathrm {obs}}$  
is the observed  radius in $arcsec$ 
and   $R_{\mathrm {ave}}$  is the averaged  
radius in $arcsec$
which is   $R_{\mathrm {ave}} = 1.54\,arcsec$.
Figure \ref{real_ring}
reports the astronomical data of SDP.81 
and the 
percentage  of
reliability is $\epsilon_{\mathrm {obs}}= 92.78\%$. 

\subsection{Canarias ER}

\label{secIAC J010127-334319}
The object  IAC J010127-334319 has been detected in the
optical region with the Gran
Telescopio CANARIAS; the radius of the ER 
is $\theta_E=2.16\,{arcsec}$, 
see \cite{Bettinelli2016}.
As an example, inserting the above  radius,
$D_{{s,{\mathrm {Mpc}}}}=1192$~Mpc,
$D_{{ls,{\mathrm {Mpc}}}}=498$~Mpc 
and $D_{{l,{\mathrm {Mpc}}}}=951$~Mpc
in Eq.~(\ref{massasun}), we obtain a mass 
for the foreground 
galaxy of 
$M (\theta_{E,arcsec}) =1.3 \,10^{12} M_{\sun}$.

\section{The equation of motion of a symmetrical SB}

\label{secmotion}
\begin{figure*}
\begin{center}
\includegraphics[width=10cm]{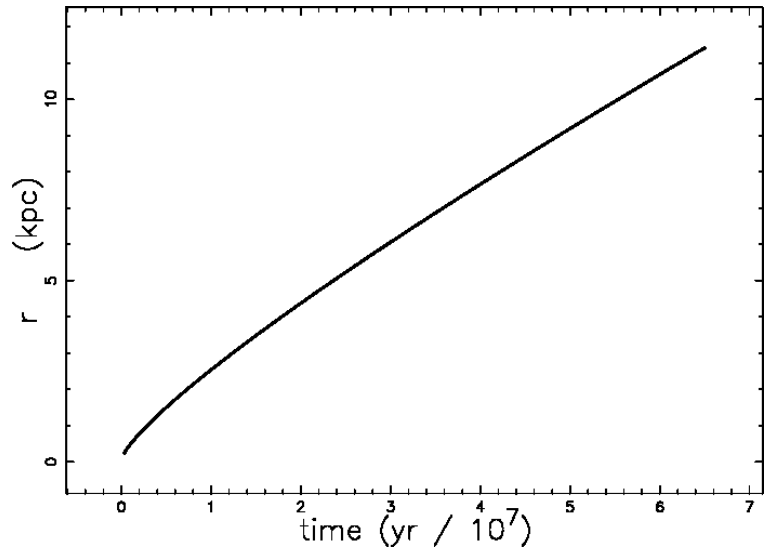}
\end {center}
\caption
{
Numerical   solution    for the radius as a function
of time for SB associated with SDP.81     (full   line), 
parameters as in Table  \ref{tablesbparameters}.
}
\label{nfw_rt}
    \end{figure*}

\begin{figure*}
\begin{center}
\includegraphics[width=10cm]{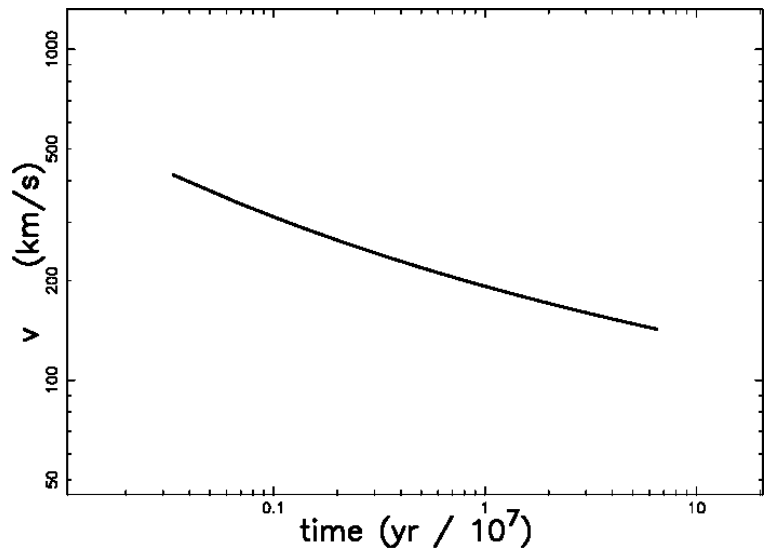}
\end {center}
\caption
{
Velocity  as a function
of time for SDP.81     (full   line), 
parameters as in Table  \ref{tablesbparameters},
both axes are logarithmic.
}
\label{nfw_vt}
    \end{figure*}

The density  is  assumed to have  a 
Navarro--Frenk--White (NFW )
dependence on $r$
in spherical  coordinates:
\begin{equation}
 \rho(r;r_0,b,\rho_0) =
\frac
             {
              \rho_{{0}}r_{{0}} \left( b+r_{{0}} \right) ^{2}
             }
             {
              r \left( b+r \right) ^{2}
             }
\quad ,
\label{profilenfw}
\end{equation}
where $b$ represents the scale, 
see \cite{Navarro1996} 
for more details.
The piece-wise  density is
\begin{equation}
 \rho (r;r_0,b,\rho_0)  = \left\{ \begin{array}{ll} 
            \rho_0                      & \mbox         {if $r \leq r_0 $ } \\
            \frac
             {
              \rho_{{0}}r_{{0}} \left( b+r_{{0}} \right) ^{2}
             }
             {
              r \left( b+r \right) ^{2}
             }
   & \mbox {if $r >    r_0 $ } 
            \end{array}
            \right.
\label{piecewisenfw}
\end{equation}
The total mass swept,   $M(r;r_0,b,\rho_0) $,
in the interval $[0,r]$ is
\begin{eqnarray}
M(r;r_0,b,\rho_0) = 
\frac {1}
{
3\,b+3\,r
}
\Bigg ( 
-4\,r_{{0}}\pi\,\rho_{{0}}  
\Big ( 3\,\ln    ( b+r_{{0}}   ) {b}
^{3}+3\,\ln    ( b+r_{{0}}   ) {b}^{2}r
\nonumber \\
+6\,\ln    ( b+r_{{0}
}   ) {b}^{2}r_{{0}}
+6\,\ln    ( b+r_{{0}}   ) brr_{{0}}
+3
\,\ln    ( b+r_{{0}}   ) b{r_{{0}}}^{2}
+3\,\ln    ( b+r_{{0}
}   ) r{r_{{0}}}^{2}
\nonumber \\
-3\,\ln    ( b+r   ) {b}^{3}-3\,\ln 
   ( b+r   ) {b}^{2}r-6\,\ln    ( b+r   ) {b}^{2}r_{{0}}
-6\,\ln    ( b+r   ) brr_{{0}}
\nonumber \\
-3\,\ln    ( b+r   ) b{r_{
{0}}}^{2}-3\,\ln    ( b+r   ) r{r_{{0}}}^{2}+3\,{b}^{2}r-3\,{b}
^{2}r_{{0}}+3\,br_{{0}}r
\nonumber  \\
-4\,b{r_{{0}}}^{2}-{r_{{0}}}^{2}r   
\Big ) 
\Bigg )
\quad .
\end{eqnarray}
The conservation of momentum in
spherical coordinates
in the framework of the thin
layer approximation  states that
\begin{equation}
M_0(r_0) \,v_0 = M(r)\,v
\quad ,
\end{equation}
where $M_0(r_0)$ and $M(r)$ are the swept 
masses at $r_0$ and $r$,
and $v_0$ and $v$ are the velocities of 
the thin layer at $r_0$ and $r$.
The velocity is, therefore, 
\begin{equation}
\frac{dr}{dt} = \frac{NE}{DE}
\quad ,
\label{vel_nfw}
\end{equation}
where
\begin{eqnarray}
NE =
-{r_{{0}}}^{2}v_{{0}} \left( b+r \right)
\nonumber  
\quad,
\end{eqnarray}
and  
\begin{eqnarray}
DE=
3\,\ln    ( b+r_{{0}}   ) {b}^{3}+3\,\ln    ( b+r_{{0}}
   ) {b}^{2}r+6\,\ln    ( b+r_{{0}}   ) {b}^{2}r_{{0}}
\nonumber  \\
+6\,
\ln    ( b+r_{{0}}   ) brr_{{0}}
+3\,\ln    ( b+r_{{0}}
   ) b{r_{{0}}}^{2}+3\,\ln    ( b+r_{{0}}   ) r{r_{{0}}}^{2
}-3\,\ln    ( b+r   ) {b}^{3}
\nonumber  \\
-3\,\ln    ( b+r   ) {b}^{2
}r-6\,\ln    ( b+r   ) {b}^{2}r_{{0}}-6\,\ln    ( b+r
   ) brr_{{0}}-3\,\ln    ( b+r   ) b{r_{{0}}}^{2}
\nonumber  \\
-3\,\ln 
   ( b+r   ) r{r_{{0}}}^{2}+3\,{b}^{2}r-3\,{b}^{2}r_{{0}}+3\,br
_{{0}}r-4\,b{r_{{0}}}^{2}-{r_{{0}}}^{2}r
\quad .
\nonumber
\end{eqnarray}
The integration of 
the above differential equation of the first order 
gives  the following non-linear equation:
\begin{eqnarray}
\frac {1}
{
{r_{{0}}}^{2}v_{{0}}
}
\Bigg (-6\,   ( b+r_{{0}}   ) ^{2}   ( b+r/2   ) \ln    ( b+
r_{{0}}   ) +6\,   ( b+r_{{0}}   ) ^{2}   ( b+r/2
   ) \ln    ( b+r   ) 
\nonumber  \\
-6\,   ( r-r_{{0}}   ) 
   ( {b}^{2}+3/2\,br_{{0}}+1/3\,{r_{{0}}}^{2}   ) \Bigg)
=\left( t-{\it t_0} \right) 
\label .
\label{eqn_nl_exp}
\end{eqnarray}
The above non-linear equation
does not have 
an analytical  
solution for the radius, $r$, 
as a function of  time.
The astrophysical  units are 
pc for length  and  yr  for time.
With these units, the initial velocity is 
$v_0$(km s$^{-1})= 9.7968 \, 10^5 v_0$(pc\ yr$^{-1})$.
The energy conserving phase of an SB 
in the presence of constant density   
allows setting up 
the initial 
conditions,
and the  radius  is 
\begin{equation}
R =  111.552\,{\frac {\sqrt [5]{{\it N^*}}\sqrt [5]{E_{{51}}}{t_{{7}
}}^{3/5}}{\sqrt [5]{n_0}}}
\quad {\mathrm{pc}} \quad ,
\label{raggioburst}
\end{equation}
where $t_7$ is the time 
expressed  in units of $10^7$ yr,
$E_{51}$  is the  energy expressed  in  units of $10^{51}$ erg,   
$n_0$ is  the number density expressed  in 
particles~$\mathrm{cm}^{-3}$
(density~$\rho_0=n_0m$, where $m=1.4m_{\mathrm {H}}$) and
$N^*$  is the number of SN explosions
in  $5.0 \cdot 10^7$ yr and therefore is a rate,
see see eq. (10.38) in \cite{McCray1987}.
The  velocity 
of an SB in such a phase is 
\begin{equation}
v_0 =  0.416324\,{\frac {{5}^{2/5}{14}^{4/5}\sqrt [5]{{\it N^*}}\sqrt 
[5]{E_{{51}}}}{\sqrt [5]{n_0}{t_{{7}}}^{2/5}}}
\quad \frac{km}{s} \quad .
\end{equation}
The initial condition for $r_0$ and $v_0$ are now fixed 
by the energy conserving phase for an SB evolving in 
a medium at constant density.
The free  parameters of the model  
are reported in Table \ref{tablesbparameters},
Figure \ref{nfw_rt} reports the law of motion and
Figure \ref{nfw_vt} the behaviour of the velocity as a function
of time.
\begin{table}[ht!]
\caption {
Theoretical  parameters of an SB evolving 
in a medium with a NFW  profile.
}
\label{tablesbparameters}
\begin{center}
\begin{tabular}{|c|c|c|c|c|c|}
\hline
Name         & $t_0$(yr)& $E_{51}$ & $n_0\,(\frac{1}{{\mathrm{cm}}^3})$   & b(pc)& $N^*$ \\
\hline                                                   
SDP.81       &  10000    &   1      & $10^{-4}$     &  1   &  1000   \\
\hline
\end{tabular}
\end{center}
\end{table}

\begin{figure*}
\begin{center}
\includegraphics[width=10cm]{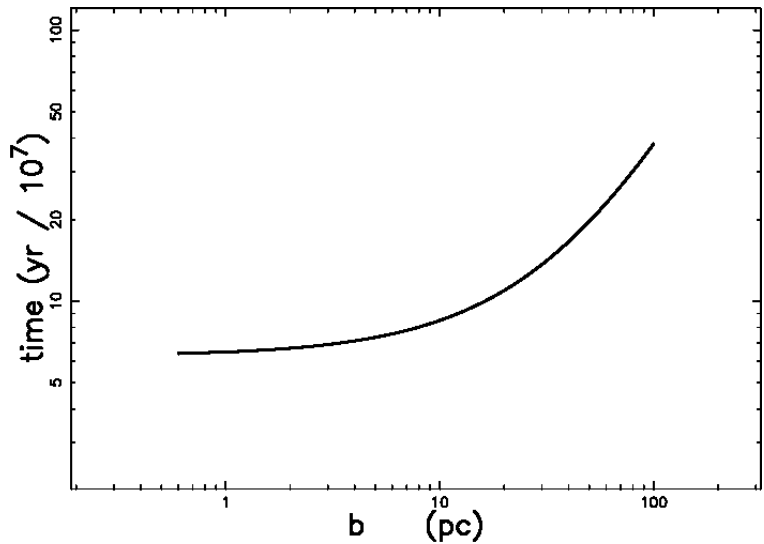}
\end {center}
\caption
{
The relation between $b$ and time 
which produces
$r=11.39$\ kpc,
other parameters as in Table  \ref{tablesbparameters}.
Both axes are logarithmic.
}
\label{nfw_bt}
    \end{figure*}

Once we have fixed the standard  radius of SDP.81 at 
$r=11.39$\ kpc, we evaluate the pair of values  
for  $b$ (the scale) and for $t$ (the time) that allows 
such a value of the radius,
see Figure \ref{nfw_bt}.

\begin{figure*}
\begin{center}
\includegraphics[width=10cm]{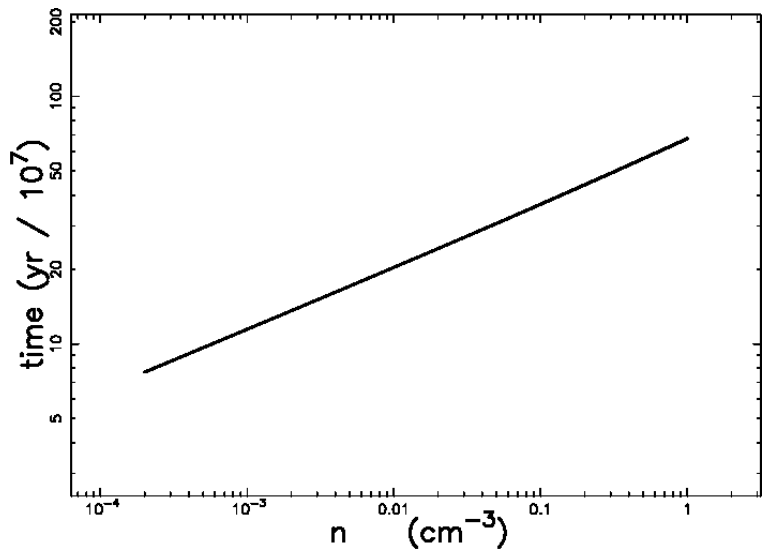}
\end {center}
\caption
{
The relation between $n_0$ and the time 
which produces
$r=11.39$\ kpc,
other parameters as in Table  \ref{tablesbparameters}.
Both axes are logarithmic.
}
\label{nfw_ant}
    \end{figure*}

\begin{figure*}
\begin{center}
\includegraphics[width=10cm]{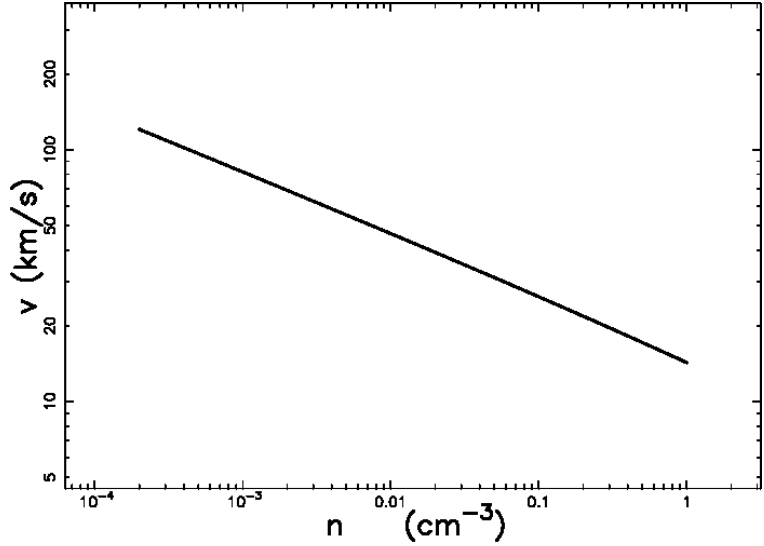}
\end {center}
\caption
{
The actual  velocity  as a function of  $n_0$ 
when the standard  radius is 
$r=11.39$\ kpc,
other parameters as in Table  \ref{tablesbparameters}.
Both axes are logarithmic.
}
\label{nfw_anv}
    \end{figure*}

The pair of values 
of $n_0$  (initial number density) and $t$ (the time) 
which produces the standard 
value of the radius is reported 
in Figure \ref{nfw_ant}; 
Figure~\ref{nfw_anv} conversely reports 
the actual velocity of the SB associated with SDP.81   as function of $n_0$. 

The swept mass can be expressed 
in  the number of solar masses, $M_{\sun}$, and, 
with parameters as in Table  \ref{tablesbparameters}, 
is
\begin{equation}
M=3732.709 \,n_0\,M_{\sun}
\quad .
\end{equation}

\section{The equation of motion of an asymmetrical SB}

\label{secmotionasymm}
In order to simulate an asymmetric SB 
we briefly review a numerical  algorithm 
developed in \cite{Zaninetti2012g}.
We assume a  number density
distribution as  
\begin{equation}
n(z) = n_0 sech^2 (\frac{z}{2\,h})
\quad ,
\label{sech2}
\end{equation}
where $n_0$ is the density at $z=0$,
$h$ is a scaling parameter, 
and  $sech$ is the hyperbolic secant  
(\cite{Spitzer1942,Rohlfs1977,Bertin2000,Padmanabhan_III_2002}).

We  now analyze the case   of an 
expansion  that starts  from a given 
galactic height $z$,  denoted by $z_{\mathrm{OB}}$,
which  represents  the OB associations.
It is not possible to find  $r$   analytically  and
a numerical method   should be implemented.

The following two recursive equations are found when
momentum conservation is applied:
\begin{eqnarray}
r_{n+1} = r_n + v_n \Delta t    \nonumber  \\
v_{n+1} = v_n 
\Bigl (\frac {M_n(r_n)}{M_{n+1} (r_{n+1})} \Bigr ) 
\quad  ,
\label{recursive}
\end{eqnarray}
where  $r_n$, $v_n$, $M_n$ are the temporary  radius,
the velocity,  and the total mass, 
respectively,
$\Delta t $ is the time step,  and $n$ is the index.
The advancing expansion is computed in a 3D Cartesian
coordinate system ($x,y,z$)  with the center 
of the explosion at  (0,0,0).
The explosion is better visualized  
in a 3D Cartesian
coordinate system ($X,Y,Z$) in which the galactic plane
is given by $Z=0$.
The following 
translation, $T_{\mathrm{OB}}$,   
relates  the two Cartesian coordinate  systems. 
\begin{equation}
T_{\mathrm{OB}} ~
 \left\{ 
  \begin {array}{l} 
  X=x  \\\noalign{\medskip}
  Y=y  \\\noalign{\medskip}
  Z=z+ z_{\mathrm{OB}}
  \end {array} 
  \right.  \quad , 
\label{ttranslation}
\end{equation}
where $z_{\mathrm{OB}}$  
is the distance  in parsec   of the 
OB associations   from the galactic plane.

The physical units for the asymmetrical SB 
have not yet been specified: 
parsecs for length and
$10^7\,yr$ for time are perhaps an acceptable 
astrophysical choice. 
With
these units, the initial velocity $v_{{0}}=\dot {r_0}$ is 
expressed in
units of pc/($10^7$ yr) and should be converted 
into km/s; this
means that $v_{{0}} =10.207 v_{{1}}$ 
where  $v_{{1}}$ is
the initial velocity expressed in km/s.

\begin{figure}
  \begin{center}
\includegraphics[width=10cm]{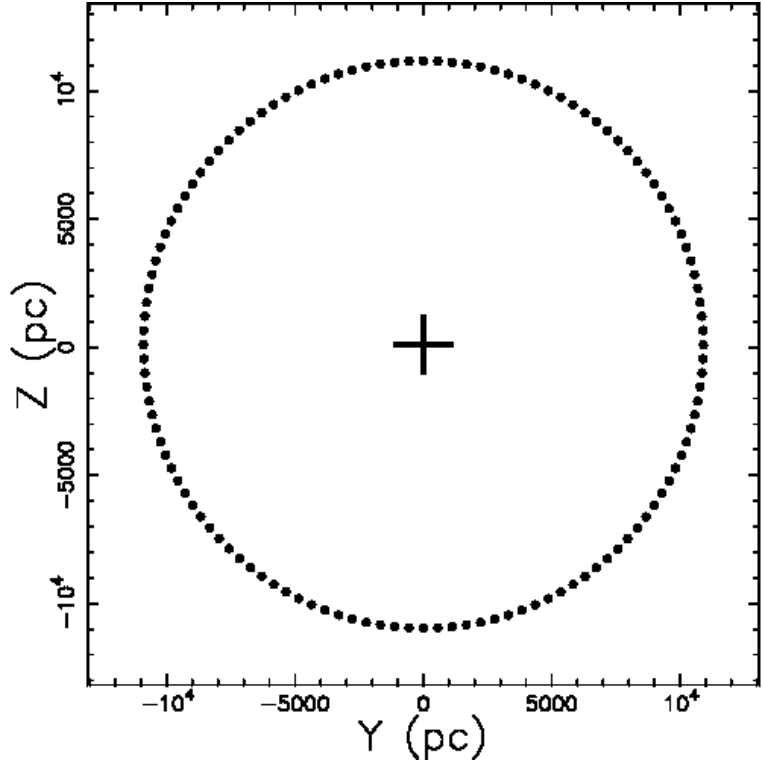}
  \end {center}
\caption
{
Section of the SB associated with SDP.81
in the {\it Y-Z;X=0}  plane
when the explosion starts at  $z_{\mathrm{OB}}=100 \mbox{pc}$.
The code parameters for  the  numerical couple  
(\ref{recursive})
are 
$h=14000$ $pc$, 
$t_7=350~$,
$t_{7,0}$ = 0.35, 
$r_0$ = 1325, 
$v_0=222 \mathrm{km}\, \mathrm{s}^{-1}$,
$N_{SN}$ = 140000  
 and  $N^*$=2000000.
The explosion site is represented by a cross.
}
\label{section_auto}%
    \end{figure}

We  are now ready to present the numerical evolution 
of the SB   associated with SDP.81
when $z_{\mathrm{OB}}=100 \mbox{pc} $,
see Fig.~\ref{section_auto}.

The degree of asymmetry can be evaluated  introducing 
the radius along 
the polar direction up,   $r_{up}$,
the polar direction down, $r_{up}$
and 
the equatorial  direction , $r_{eq}$.
In our model all the already defined three radii 
are different,
see Table \ref{tabdirections}.

\begin{figure}
\includegraphics[width=10cm]{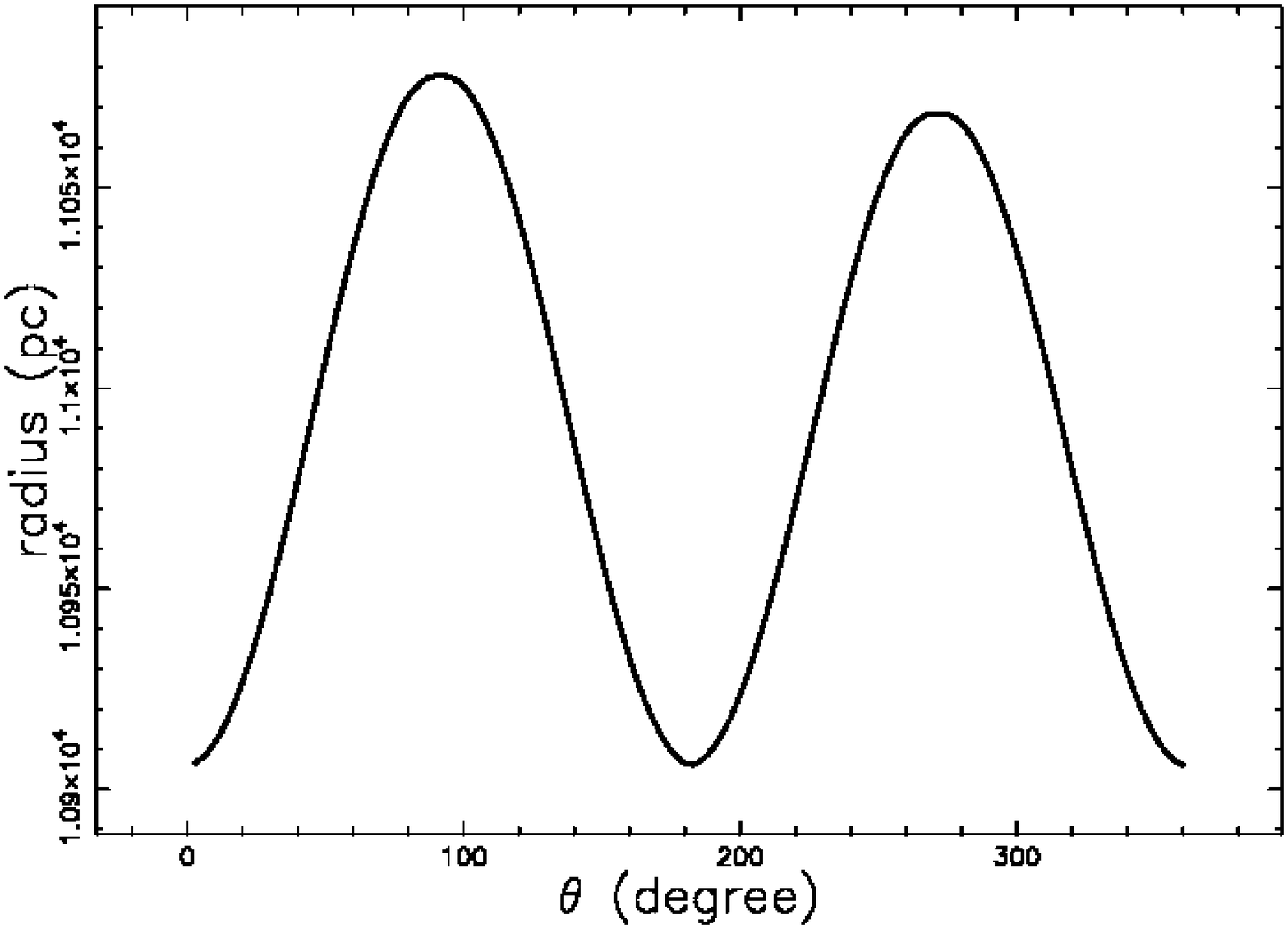}
\caption 
{
Radius in $pc$  of the SB associated with SDP.81 as a function of
the position   angle   in degrees for
a self-gravitating  medium,
parameters as in Figure \ref{section_auto}.
}%
    \label{radius_theta}
    \end{figure}

\begin{figure}
\includegraphics[width=10cm]{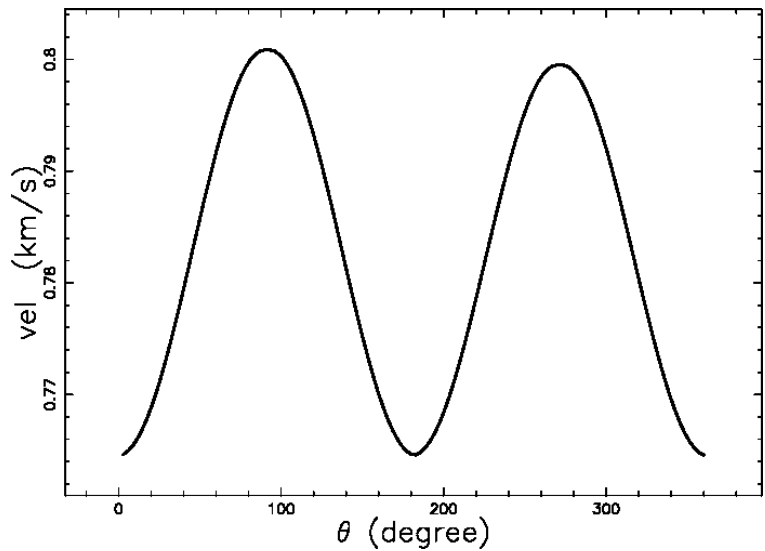}
\caption 
{
Velocity  in $\frac{km}{s}$  of the the SB associated 
with SDP.81 as a
function of the position   angle in degrees
for a self-gravitating  medium,
parameters as in Figure \ref{section_auto}.
}%
    \label{vel_theta}
    \end{figure}

   \begin{table}
      \caption{Radii concerning  SB associated with SDP.81,
               parameters as in Figure \ref{section_auto}. }
         \label{tabdirections}
      \[
         \begin{array}{cc}
            \hline
            \noalign{\smallskip}
\mathrm{Direction}& r  (\mathrm{pc}) \\
            \noalign{\smallskip}
            \hline
            \noalign{\smallskip}
\mathrm{Equatorial} & 10906   \\
\mathrm{Polar~up}   & 11077   \\
\mathrm{Polar~down} & 11068   \\
            \noalign{\smallskip}
            \hline
         \end{array}
      \]
   \end{table}
We can evaluate  the radius  and the velocity
as function of the direction 
plotting the radius and the direction
in section in the {\it Y-Z;X=0}  plane,
see Figures 
\ref{radius_theta} 
and 
\ref{vel_theta}.   

\section{The  image}

We now  briefly review  the basic equations of the radiative transfer
equation, the conversion of the flux of energy 
into luminosity, the 
symmetric and the asymmetric theory of the image.

\subsection{Radiative transfer equation}
\label{sec_transfer}
The transfer equation in the presence of emission and absorption, see for
example eqn.(1.23) in \cite{rybicki} 
or eqn.(9.4)       in \cite{Hjellming1988} 
or eqn.(2.27)      in \cite{Condon2016},
 is
 \begin{equation}
\frac {dI_{\nu}}{ds} =  -k_{\nu} \zeta I_{\nu}  + j_{\nu} \zeta
\label{equazionetrasfer} \quad ,
\end {equation}
where  $I_{\nu}$ is the specific intensity or spectral brightness, 
$s$  is the line of
sight, $j_{\nu}$ the emission coefficient, $k_{\nu}$   a mass
absorption coefficient, $\zeta$ the  mass density at position $s$,
and the index $\nu$ denotes the involved frequency of emission.
The solution to  Eq.~(\ref{equazionetrasfer})
 is
\begin{equation}
 I_{\nu} (\tau_{\nu}) =
\frac {j_{\nu}}{k_{\nu}} ( 1 - e ^{-\tau_{\nu}(s)} ) \quad  ,
\label{eqn_transfer}
\end {equation}
where $\tau_{\nu}$ is the optical depth at frequency $\nu$
\begin{equation}
d \tau_{\nu} = k_{\nu} \zeta ds \quad.
\end {equation}
We now continue analysing the case of an
 optically thin layer
in which $\tau_{\nu}$ is very small (or $k_{\nu}$  very small)
and the density  $\zeta$ is replaced by the  number density
of  particles, $n(s)$.
 In the following, the 
emissivity is taken to be proportional to the number density 
\begin{equation}
j_{\nu} \zeta =K  n(s) \quad  ,
\end{equation}
where $K$ is a  constant.
The intensity is therefore
\begin{equation}
 I_{\nu} (s) = I_0 + K
\int_{s_0}^s   n (s\prime) ds \prime ,
\label{transport}
\end {equation}
where $I_0$ is the intensity 
at the  point $s_0$.
The MKS units of the intensity are W\ m$^{-2}$\ Hz${^-1}$ \ sr$^{-1}$.
The increase in brightness
is proportional to the number density 
integrated along
the line of  sight: in the case of constant number density,
it is proportional only to the line of  sight.

As an example, synchrotron emission
has an intensity proportional to $l$, the dimension 
of the radiating region, in the case of a constant 
number density of the radiating particles, 
see formula (1.175) of \cite{lang}. 

\subsection{The source of luminosity}

The ultimate source of the  observed luminosity
is assumed  to be
the rate of kinetic energy,
$L_m$,
\begin{equation}
L_m = \frac{1}{2}\rho A  V^3
\quad,
\label{fluxkineticenergy}
\end{equation}
where $A$ is  the considered area,
      $V$ is  the velocity of a spherical SB
and   $\rho$  is the density in the advancing layer of
a spherical SB.
In the case of the spherical expansion of an SB, $A=4\pi r^2$,
where $r$ is the instantaneous radius of the SB,
which means
\begin{equation}
L_m = \frac{1}{2}\rho 4\pi r^2 V^3
\quad .
\label{fluxkinetic}
\end{equation}
The  units of the luminosity  are W in MKS 
and erg\ s$^{-1}$ in CGS.
The astrophysical  version of the 
the rate of kinetic energy, $L_{ma}$, 
is 
\begin{equation}
L_{ma} =
{ 1.39\times 10^{29}}\,{\it n_1}\,{{\it r_1}}^{2}{{\it v_1}}^{3}
\frac{{\mathrm{ergs}}}{{\mathrm{s}}}
\quad,
\label{kineticfluxastro}
\end{equation}
where $n_1$   is the   number density expressed
in units  of  $1~\frac{{\mathrm{particle}}}{{\mathrm{cm}}^3}$,
$r_1$  is  the  radius in parsecs,
and
$v_{1}$ is the   velocity in
km/s.
As an example, according to Figure \ref{nfw_anv},
inserting 
$r_1=11.39\,10^3$, 
$n_1=0.1$ and 
$v_{1}$ =26.08 
in the above formula, 
the maximum available mechanical luminosity is 
$L_{ma} =3.2\,10^{40}
\,\frac{{\mathrm{ergs}}}{{\mathrm{s}}}$.  
The spectral luminosity, $ L_{\nu} $,
at a given frequency $\nu$
is
\begin{equation}
L_{\nu} =  4 \pi  \dl^2  S_{\nu}
\quad ,
\label{dasalum}
\end{equation}
where   $S_{\nu}$ is the observed flux density 
at a given frequency $\nu$
with MKS units as W\ m$^{-2}$\ Hz$^{-1}$.
The   observed
luminosity 
at a given frequency $\nu$
can  be expressed as
\begin{equation}
L_{\nu} = \epsilon  L_{ma}
\label{luminosity}
\quad ,
\end{equation}
where  $\epsilon$  is  a conversion constant
from  the mechanical luminosity   to  the
observed luminosity.
More details  on the synchrotron luminosity 
and the connected astrophysical units 
can be found in \cite{Condon2016}.

\subsection{The symmetrical image theory}

\label{ringsec}
\begin{figure*}
\begin{center}
\includegraphics[width=10cm]{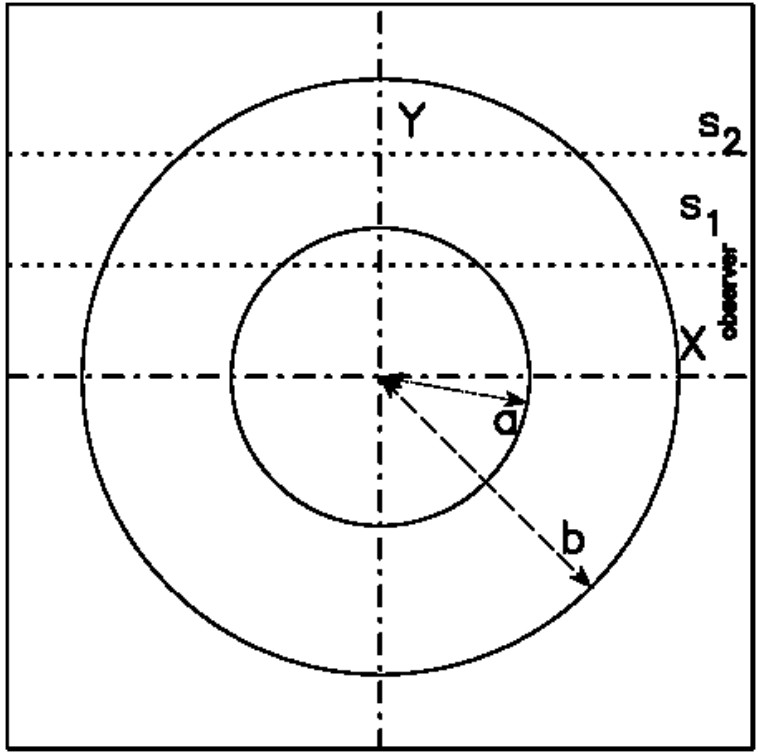}
\end {center}
\caption
{
The two circles (sections of spheres)  which   include the region
with constant number density of emitting matter
are   represented by
a full line.
The observer is situated along the $x$ direction, and 
three lines of sight are indicated.
}
\label{plotab}
    \end{figure*}

\begin{figure*}
\begin{center}
\includegraphics[width=10cm]{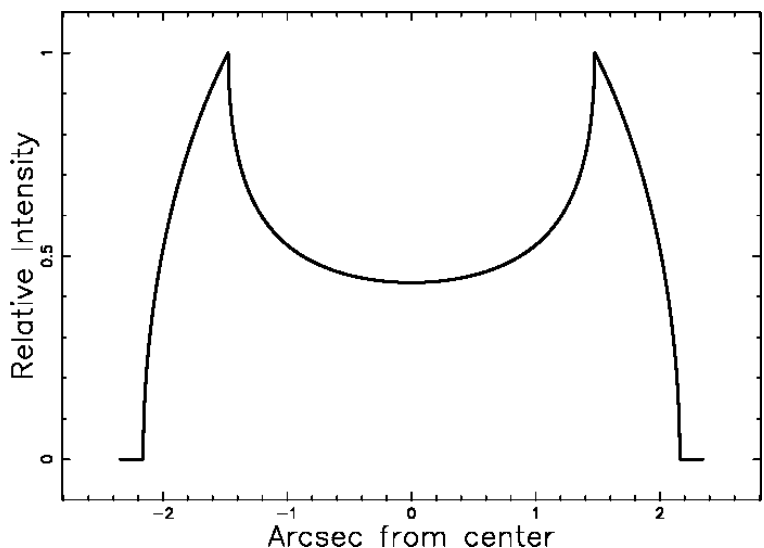}
\end {center}
\caption
{
 Cut of the intensity ${\it I}$
 of the ring  model, Equation~(\ref{irim}), 
 crossing the center. 
 The $x$ and  $y$ axes  are in 
 $arcsec$, 
 $a=1.23\,arcsec$,   
 $b=1.54\,arcsec$
 and $\frac {I(y=a)} {I(y=0)}=3$.
 This cut refers to SDP.81.
}
\label{ring_cut}
    \end{figure*}

\begin{figure*}
\begin{center}
\includegraphics[width=10cm]{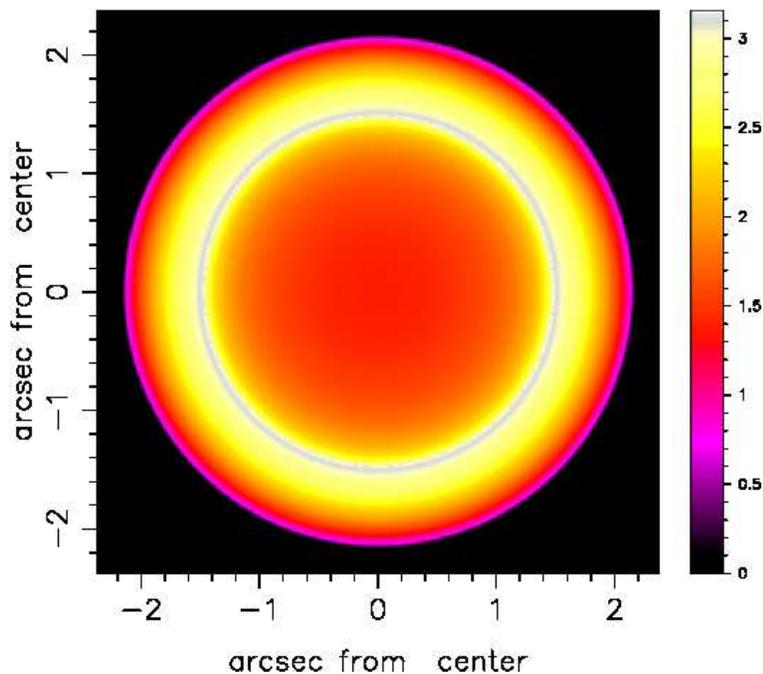}
\end {center}
\caption
{
Contour map  of  ${\it I}$,
the $x$ and  $y$  axes  are in $arcsec$,
parameters as in Figure \ref{ring_cut}.
}
\label{circle}
    \end{figure*}

We assume that the number density 
of the emitting matter    
$n$ is variable, and in particular
rises from 0 at $r=a$ to a maximum value $n_m$, remains
constant  up to $r=b$, and then falls again to 0.
This geometrical  description is shown in  
Figure~\ref{plotab}.
The length of the line of sight, when the observer is situated
at the infinity of the $x$-axis, 
is the locus    
parallel to the $x$-axis which  crosses  the position $y$ in a 
Cartesian $x-y$ plane and terminates  at the external circle
of radius $b$.
The locus length is   
\begin{eqnarray}
l_{0a} = 2 \times ( \sqrt { b^2 -y^2} - \sqrt {a^2 -y^2}) 
\quad  ;   0 \leq y < a  \nonumber  \\
l_{ab} = 2 \times ( \sqrt { b^2 -y^2})  
 \quad  ;  a \leq y < b    \quad . 
\label{length}
\end{eqnarray}
When the number density
of the emitting matter 
 $n_m$ is constant between two spheres
of radii $a$ and $b$,
the intensity of radiation is 
\begin{eqnarray}
I_{0a} =K_I \times n_m \times 2 \times ( \sqrt { b^2 -y^2} - \sqrt {a^2 -y^2}) 
\quad  ;   0 \leq y < a  \nonumber  \\
I_{ab} =K_I \times n_m \times  2 \times ( \sqrt { b^2 -y^2})  
 \quad  ;  a \leq y < b    \quad , 
\label{irim}
\end{eqnarray}
where $K_I$ is a constant.
The ratio between the theoretical intensity at the maximum $(y=a)$
 and at the minimum ($y=0$)
is given by 
\begin{equation}
\frac {I(y=a)} {I(y=0)} = \frac {\sqrt {b^2 -a^2}} {b-a}
\quad .
\label{ratioteorrim}
\end{equation}
The  parameter $b$ is identified with the external radius, which means 
the advancing radius of an SB.
The parameter $a$ can be found from 
the following formula:
\begin{equation}
a  = \frac
{
b \left(   (\frac {I(y=a)} {I(y=0)})_{obs}^2 - 1  \right) 
}
{
\left(   (\frac {I(y=a)} {I(y=0)})_{obs}^2 + 1  \right) }
\quad ,
\end{equation}
where  $(\frac {I(y=a)} {I(y=0)})_{obs} $ 
is the observed ratio between 
the maximum intensity at  the rim 
and the intensity at the center.
The distance $\Delta y$ after which the intensity 
is decreased of a factor $f$ 
in the region $a \leq y < b$  
is  
\begin{equation}
\Delta y =
\frac{2   \sqrt {{b}^{2}{f}^{2}+{a}^{2}-{b}^{2}}
-\sqrt {{a}^{2}{f}^{4}-{b}^{2
}{f}^{4}+2   {a}^{2}{f}^{2}+2   {b}^{2}{f}^{2}+{a}^{2}-{b}^{2}}}{2 \,f} 
\quad .
\end{equation}
We can now evaluate the half-width half-maximum
by analogy with the Gaussian profile
 $HWHM_U$,
which  is obtained by the previous formula
upon inserting $f=2$:
\begin{equation}
HWHM_U = 
\frac{1}{2}\,\sqrt {{a}^{2}+3\,{b}^{2}}-\frac{1}{4}\,\sqrt {25\,{a}^{2}-9\,{b}^{2}}
\quad .   
\end{equation}
In the above model, $b$ is associated with 
the radius of the 
outer  region of the observed ring, $a$ conversely 
can be deduced from  the observed 
$HWHM_U$: 
\begin{eqnarray}
a = \nonumber \\
\frac{1}{21}\,\sqrt {441\,{b}^{2}+464\,{{\it HWHM_U}}^{2}-32\,\sqrt {441\,{b}^
{2}{{\it HWHM_U}}^{2}+100\,{{\it HWHM_U}}^{4}}}
\quad  .
\end{eqnarray}
As an example, inserting in the above formula 
$b=1.54\,arcsec$   and   $HWHM_U=0.1\,arcsec $, we 
obtain $a=1.46\,arcsec$. 
A cut in the theoretical intensity 
of SDP.81, see Section \ref{secsdp81},   
is reported in Figure~\ref{ring_cut}
and  a  theoretical   image in Figure \ref{circle}. 


\begin{figure*}
\begin{center} 
\includegraphics[width=10cm]{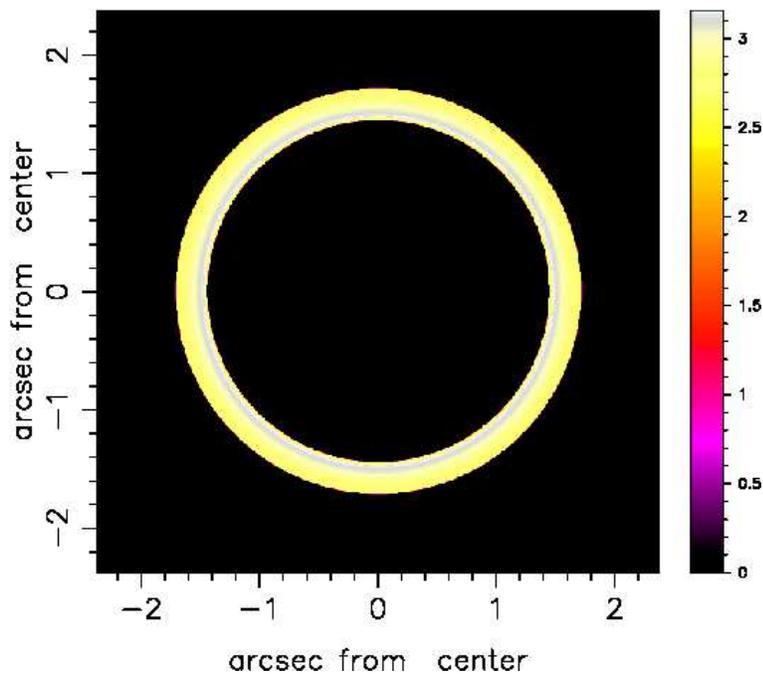}
\end {center}
\caption {   
 The same  as  Figure \ref{circle}  but with
 $I_{tr}= I_{max}/fac$,
parameters as in Figure \ref{ring_cut} and $fac=1.2$.
} \label{circle_hole}
    \end{figure*}

The effect of the  insertion of a threshold 
intensity, $I_{tr}$,
which is connected with the observational 
techniques, is now analysed. 
The
threshold intensity can be parametrized  to  $I_{max}$, 
the 
maximum  value  of intensity characterizing the ring: 
a typical
image with  a hole  is visible in  
Figure~\ref{circle_hole} when
$I_{tr}= I_{max}/fac$, where $fac$ is a parameter
which allows matching theory with observations.
A comparison between the theoretical 
intensity  and the theoretical flux 
can be made through the formula
(\ref{dasalum})  
and due to the  fact that $\dl$ is assumed to constant 
over all the astrophysical image, 
the theoretical intensity  and the theoretical flux  are
assumed to scale in the same way.

\begin{figure*}
\begin{center}
\includegraphics[width=10cm]{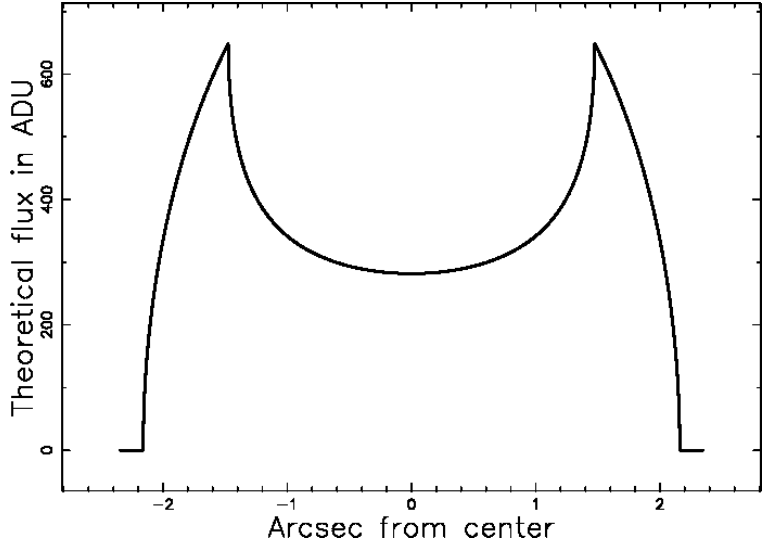}
\end {center}
\caption
{
 Theoretical flux  for the Canarias ring. 
 The $x$  and  $y$ axes  are in 
 $arcsec$, 
 $a=1.47\,arcsec$,   
 $b=2.16\,arcsec$
 and $\frac {I(y=a)} {I(y=0)}=2.30$.
 This cut refers to  IAC J010127-334319.
}
\label{ring_cut_canarian}
    \end{figure*}

The  theoretical flux profiles 
for 
IAC J010127-334319, see Section 
\ref{secIAC J010127-334319}, 
is reported in Figure  \ref{ring_cut_canarian}.

The linear relation between the angular distance, 
in pc,
and the projected distance on the sky in $arcsec$
allows to state  the  following

\begin{mydeftheorem}
The `U' profile  of cut in theoretical flux for a symmetric  ER is independent
of the exact value of the angular distance.
\end{mydeftheorem}

\subsection{The asymmetrical image theory}

\begin{figure}
\includegraphics[width=10cm]{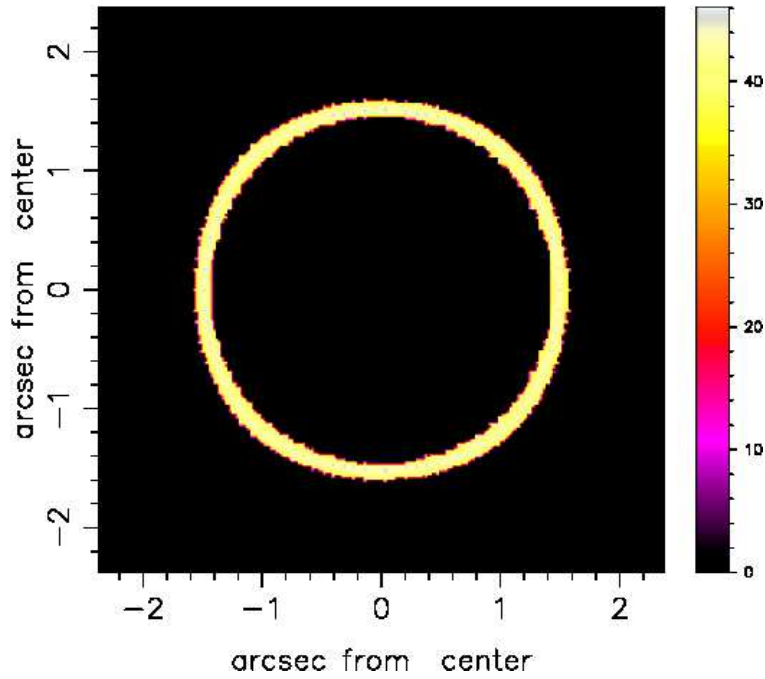}
\caption {
Map of the theoretical intensity  of
the Canarias ring.
Physical parameters as in Fig. \ref{section_auto}.
The three Euler angles
characterizing the   orientation
  are $ \Phi  $=90 $^{\circ }$,
      $ \Theta$=90 $^{\circ }$
and   $ \Psi  $=90 $^{\circ }$.
This  combination of Euler angles corresponds
to the rotated image with the polar axis along the
z-axis.
In this map $I_{tr}= I_{max}/1.2 .$
}%
    \label{ring_asym_hole}
    \end{figure}

We now explain  
a  numerical algorithm which allows us to
build  the  complex  image  
of an asymmetrical SB.
\begin{itemize}
\item 
An empty (value=0)
memory grid  ${\mathcal {M}} (i,j,k)$ which  contains
$NDIM^3$ pixels is considered
\item 
We  first  generate an
internal 3D surface by rotating the section of 
 $180^{\circ}$
around the polar direction and 
a second  external  surface at a
fixed distance $\Delta R$ from the first surface. 
As an example,
we fixed $\Delta r$ = $ 0.03 r_{max}$, 
where $r_{max}$ is the
maximum radius of expansion.
The points on
the memory grid which lie between the 
internal and the external
surfaces are memorized on
${\mathcal {M}} (i,j,k)$ with a variable integer
number   according to formula
(\ref{fluxkinetic})  and   density $\rho$ proportional
to the swept    mass.
\item Each point of
${\mathcal {M}} (i,j,k)$  has spatial coordinates $x,y,z$ 
which  can be
represented by the following $1 \times 3$  matrix, $A$,
\begin{equation}
A=
 \left[ \begin {array}{c} x \\\noalign{\medskip}y\\\noalign{\medskip}{
\it z}\end {array} \right]
\quad  .
\end{equation}
The orientation  of the object is characterized by
 the
Euler angles $(\Phi, \Theta, \Psi)$
and  therefore  by a total
$3 \times 3$  rotation matrix,
$E$, see \cite{Goldstein2002}.
The matrix point  is
represented by the following $1 \times 3$  matrix, $B$,
\begin{equation}
B = E \cdot A
\quad .
\end{equation}
\item
The intensity map is obtained by summing the points of the
rotated images
along a particular direction.
\item
The effect of the  insertion of a threshold intensity, $I_{tr}$,
given by the observational techniques,
is now analysed.
The threshold intensity can be
parametrized by $I_{max}$,
the maximum  value  of intensity
which  characterizes  the map,
see  \cite{Zaninetti2012b}. 
\end{itemize}
An  ideal image of the intensity of the Canarias ring
is shown in Fig. \ref{ring_asym_hole}.

\begin{figure}
\includegraphics[width=10cm]{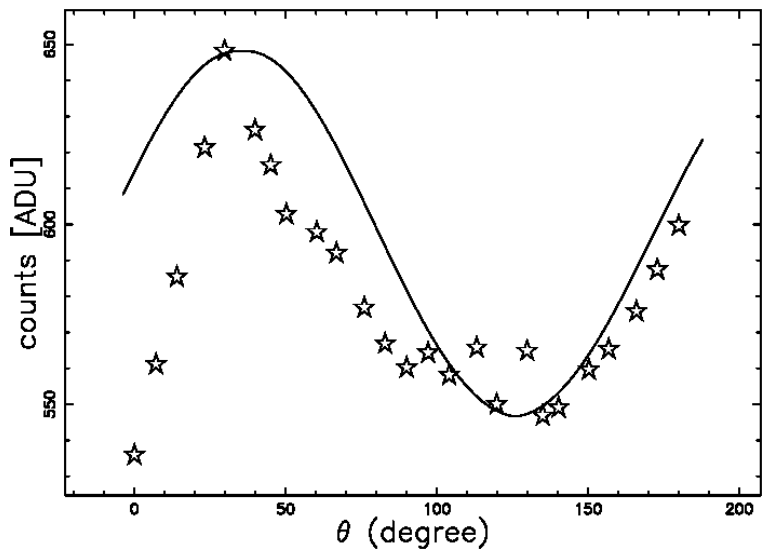}
\caption 
{
Theoretical  counts (full line  ) 
and observed counts (empty stars)
in ADU  for  the SB associated with SDP.81 
as a function of
the position   angle,   
parameters as in Figure \ref{section_auto}.
}%
    \label{flux_theo_obs_theta}
    \end{figure}

The theoretical  flux  which  is here assumed to scale
as  the flux  of  kinetic  energy  
as represented by  eqn.(\ref{fluxkinetic}),
is reported in Figure \ref{flux_theo_obs_theta}.
The 
percentage  of
reliability 
which characterizes  the observed and  
the theoretical variations in intensity
of the above figure  
is $\epsilon_{\mathrm {obs}}= 92.7\%$.

\section{Conclusions}

{\bf Flat cosmology:}
In order to have  a reliable evaluation 
of the radius of SDP.81 we have  provided 
a Taylor approximation of order 10 
for the  luminosity distance in the framework of
the flat cosmology.
The  percentage error
between analytical solution and approximated solution 
when  $z=3.04$ (the redshift of SDP.81)
is $\delta = 0.588\%$. 

{\bf Symmetric evolution of an SB:}
The motion of a SB advancing in a medium with 
decreasing density in spherical symmetry 
is analyzed.
The type of density profile here adopted is
a NFW  profile which has three free parameters,
$r_0$ ,$b$  and $\rho_0$.
The available astronomical data does not allow
to close the equations at  
$r=11.39$\ kpc (the radius of SDP.81).
A numerical relationship which connects the number 
density  with the lifetime of a SB is
reported in Figure \ref{nfw_ant}
and an approximation of the above relationship  is
$\frac{t}{10^7\,yr} = 67.36 n_0^{0.26}$ when
$b=1\,pc$  and  $r_0=44 \, pc$.

{\bf Symmetric Image theory:} 
The transfer equation for the luminous intensity
in the case of optically thin layer 
is reduced in the case of spherical symmetry 
to the evaluation of a  length between
lower and upper radius
 along 
the line of sight, see eqn.\ref{length}.
The cut in intensity has a characteristic "U" shape, see
eqn.(\ref{irim}), 
which also characterizes  the image of ER associated 
with the galaxy SDP.81.

{\bf Asymmetric Image theory:} 

The layer between  a complex 3D advancing surface 
with radius, $r_a$,  function of two angles in polar 
coordinates (external surface) and $r_a- \Delta r$ 
(internal surface) is filled with  N random points.
After a rotation characterized  by three Euler 
angles which align the 3D layer 
with the observer, the image is obtained
by summing a 3D visitation grid 
over one index, see image \ref{ring_asym_hole}.
The variations 
for the  Canarias ER
of the flux counts in ADU
as  function of the angle
can be modeled because radius, velocity and 
therefore flux of kinetic energy
are different for each chosen direction,
see Figure \ref{flux_theo_obs_theta}.

\section*{Acknowledgments}
The real data of  Figure \ref{flux_theo_obs_theta}
were kindly provided by
Margherita Bettinelli.
The real data of Figure \ref{real_ring} were digitized 
using  WebPlotDigitizer, a
Web based tool to extract data from plots,
available at  \url{http://arohatgi.info/WebPlotDigitizer/}.

\noindent
{\bf REFERENCES}

\providecommand{\newblock}{}

\end{document}